## Bringing Physical Physics Classroom Online – Challenges of Online Teaching in the New Normal

Da Yang Tan & Jer-Ming Chen

Science, Mathematics and Technology, Singapore University of Technology and Design

The onset of the COVID-19 pandemic in 2020 has greatly impacted all forms of social activities globally, including traditional classroom activities across all levels of instruction (kindergarten to universities). While many countries have opted for suspension of lessons, this cannot continue indefinitely and alternative means to continue lessons must be developed. While online and blended learning (including MOOCs) have been an active subject of research and discourse during the pre-pandemic days[1,2,3,4,5], onset of the pandemic has suddenly created an immediacy to such means of course delivery[6,7,8,9] better than any administrator or teaching committee could have done. This creates both a gap and tension in terms of successful and engaging content delivery, where traditional modes of synchronous content delivery is now forced to be brought online. Yet, the situation provides educators with an opportunity to explore the merits and weaknesses of online learning. Thus, this article seeks to outline the challenges and paradigm shifts involved in such synchronous online learning as a replacement for traditional classroom learning, following our experience at SUTD of conducting a full 13-week online physics course between May to August 2020. At the same time, we reflect on the merits brought about by the availability of such technologies that can potentially be translated to the physical physics classrooms.

**Context**

Physical World[10] is a compulsory introductory freshman undergraduate physics course that covers both classical mechanics and thermodynamics, for students preparing to be engineers or architects. Due to the special circumstance brought by the pandemic outbreak, the course was offered as a special summer course for undergraduates who matriculated a term earlier than the regular term and was brought fully online for the first time. The regular physical course is conducted in a classroom setting, with each class typically consisting of about 50 students and two instructors[11], and two weekly session lasting 2.5 hours each time. Each session covers a set of core contents, and students are typically asked to collaboratively solve two to three 'case problems' pertaining to the content at suitable time points during the class in groups of 4 to 5. For this special summer course, the session was scaled down to 34 students, but was helmed by two instructors, who would take turns to lead the session.

In light of the active-learning paradigm adopted by SUTD's undergraduate pedagogy plan, the left figure in Figure 1 shows the typical setup in the Physical World class, which aims to be both interactive and collaborative. In a typical class, students will be engaged in in-class activities, such as various hands-on activities (e.g. using 2D kinematics to launch a projectile to a target and measuring the moment of inertia of a rotating bicycle wheel attached to a mass) and solving of case problems. The instructors also have a set repertoire of physics demonstrations in many of the physical lessons. Therefore, a key challenge[12] would be to develop means for such collaborative activities to take place in an online environment, so as to maintain active-learning and to avoid a 'teacher-as-a-talking-head' scenario common in online teaching. As such, we will focus on the in-class online experience afforded to the students below. We highlight three of the challenges that we encountered and suggest potential mitigations.

**Challenge 1: Replication of the Collaborative Learning within the Online Classroom**

The first challenge would be to, as far as possible, replicate the similar form of collaborative learning experience afforded by a physical classroom. We attempted to mimic the collaborative case problems solving via the 'breakout room' function on a video conferencing platform, see Figure 1. Students were divided into groups of four or five during discussions and solving of in-class case problems. The use of breakout rooms (right side of Figure 1) proved to be useful as students were able to solve questions collaboratively, using the whiteboard feature to annotate their solutions. Such ability is essential for an equations-heavy subject like physics, since regular typesetting would be cumbersome and impede real-time discussions. However, in practice, that demanded the students to have an appropriate set-up, such as a stylus-enabled laptop or writing device, which may not be available to all.

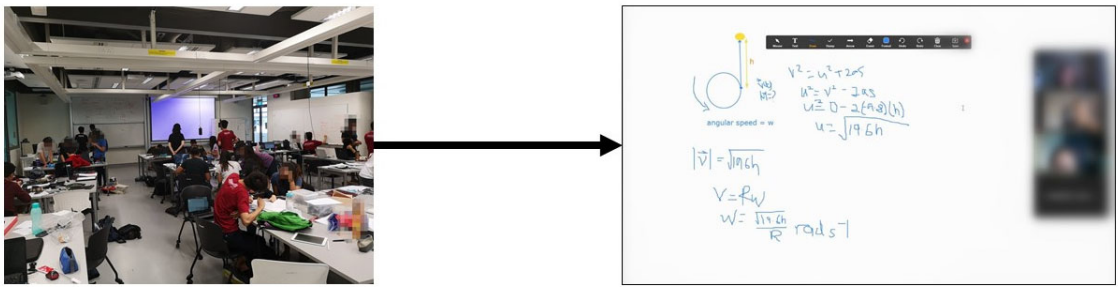

Figure 1 The transition from collaborative learning in a physical classroom to an online 'breakout room'. A typical classroom discussion is shown in the left picture. The right picture shows the screenshot of a group of four students solving a kinematics case problem during the lesson.

We demonstrate the tension in Figure 2, which shows one such artefact from the students' discussion of the in-class case problems. The artefact shows a distinctive differing affordance of technological resources within the students in the group. One of the students had access to writing devices and was able to directly 'doodle' on the virtual whiteboard, while the other two students were only able to type or to use standard shapes to draw diagrams. In our observations, this slowed down discussions as students struggled to type out the equations and hence extra time was had to be given for the students to complete their discussions.

A possible partial mitigation would be to ensure that at least one student within the group have access to such a device and assigned as the group recorder, although we noted some group members still opt to type out their working in order to communicate their ideas, which was what likely happened in Figure 2. We note however, by the end of the course, some students had become quite adept at using in-built geometric shapes to build and annotate 'sketches' and equations.

Nonetheless, we found that such collaborative discussions could take place productively, albeit at a slower pace. More importantly, instructors were able to go to the different breakout rooms to observe the ongoing discussions, like how the instructors went from group to group in a physical classroom.

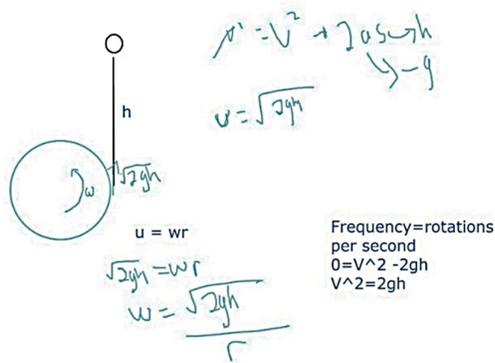

Figure 2 In solving of a kinematics case problem, students were divided into breakout rooms virtually. This artefact highlights the difference in the technology tools afforded by different students within the breakout room.

**Challenge 2: Physics Demonstration**

No physics lesson can be completed without live demonstrations and activities connecting theory with real world experiences. Unfortunately, the online mode of instruction drastically limits the wonder and visual impact to the students steered by such live demonstrations. We suggest carefully curating demonstrations that fit easily into the field of view of the camera (and picked up by the microphone, if sound is involved), without having the instructor to move about significantly, given that the instructor may similarly be constrained by his own teaching space. Here we provide three such examples in Figure 3:

Figure 3(a) shows a demonstration within the topic of forces, where we challenged the students to consider the amount of force that the atmosphere exerted on us. By using everyday objects, such as a silicone or rubber lids, to hold up a heavy tablet, one could appreciate the actual amount magnitude of force that we are subjected to everyday from our atmosphere. Figure 3(b) and 3(c) are commercially available demonstrations that were used when teaching resonance and heat engine respectively. Importantly, we quickly realized that *tabletop demonstrations* worked the best, as they could fit easily within the field of view of the [laptop] camera[13]. Another consideration as exhibited by these demonstrations are that one should choose demonstrations with non-intuitive outcomes are not expected by the students. This create an element of surprise, which then makes up for the lack of drama and suspense afforded by a live demonstration experience.

On the other hand, one should also consider experiments that are mainly visual rather than auditory. In the resonance demonstration, the perceived effect of the resonant sound produced by the second passive sound box was largely limited by the capability of the microphone receiver on the instructor's end and the speaker/headphone at the students' end: some students flagged that they were unable to hear clearly when the resonances were mismatched (a subtler sound), and this limited the expected learning outcome of this demonstration.

**Challenge 3: Non-verbal Feedback and Maintaining Student Engagement Online**

Obtaining non-verbal feedback via body-language cues from students during the lesson is one of the most important mechanism needed to enable instructors to adjust and optimize their lessons accordingly. However, much of this information is lost during an online set-up. For example, since any sound that is made through the microphone will be amplified greatly during the session and disrupt the class completely, therefore only instructor will have the microphone unmuted for most duration of the lesson. As such, some of the audio cues will be lost, for instance, the sigh of relief or the eyes 'lighting up' when the student finally understands a complicated concept. While visual feedback can be partially maintained via the instructors' insistence that the students' web cameras be kept switched on for the entire lesson[14], the cognitive load on the instructors nonetheless increases significantly as the instructor will now not only have to deliver the lesson, but also to scroll through the video and chat pages to see how the students are coping at the same time. While there is no easy solution to the increased cognitive load, our experiences indicated that the use of multiple monitors would relieve the load slightly[15]. With dual monitors, one of which may be used to display the students' video streams, while second monitor could be used to display the instructor's own video stream (or presentation slides/whiteboard). On the other hand, instructors may consider providing more breaks in between lessons[16], or reduce the duration of the lesson. Having a second instructor available to assist and address questions and provide additional feedback to the main instructor is also helpful.

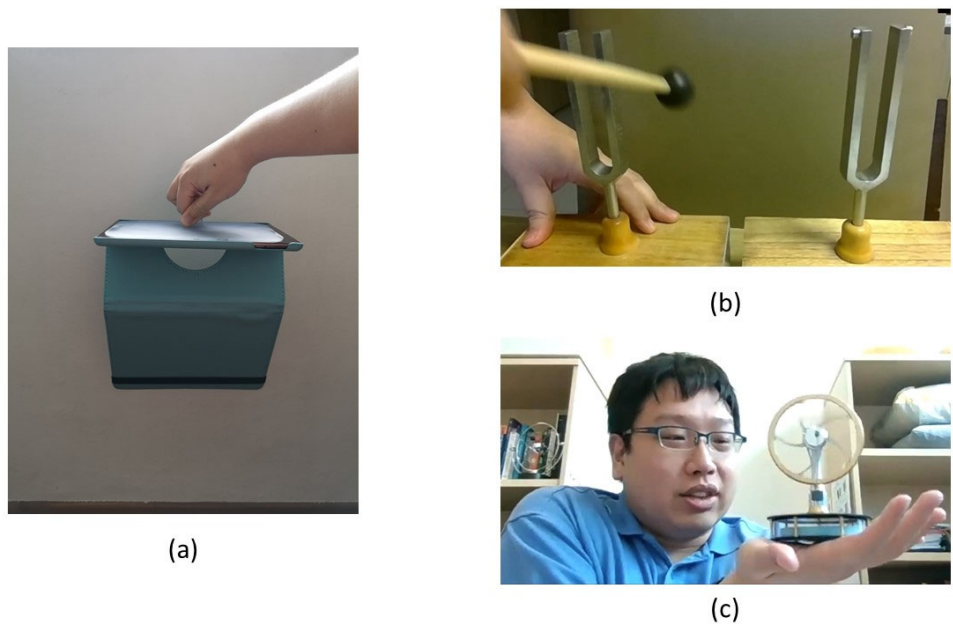

Figure 3 Demonstrations that were curated for our online lessons: (a) Using a rectangular silicone lid to demonstrate the amount of force that the atmosphere can exert on a large surface. The silicone lid was able to hold up a tablet in the air; (b) Demonstrating resonance using two 440 Hz tuning fork, each attached to a sound box, and one driving the other; (c) Showcasing the Stirling engine using a low temperature differential engine with the temperature difference provided by the warmth of the hand and the surrounding environment at about 20 degrees Celsius.

On the other hand, the chat feature within the video conferencing platform[20] provides an avenue for live feedback and engagement between the instructors and learners. Here we recommend instructors to regularly check the chat box and address the questions frequently and while the topic is 'hot', this will provide for a positive feedback loop where students know that their questions will be addressed quickly, thereby encouraging participation. Another use for the live chat would be to solicit reflections. In our classes, we asked students a number of questions at the end of each week, such as "Name a concept that inspires you this week" and "What is the most important concept that you have learnt" etc. Notice that these questions aim to engage the students by invoking emotions, targeting their affective domain. Several video conferencing platforms also provide additional features such as polls, and these can be used strategically by the instructors to engage the students further, as shown in Figure 4.

Bringing synchronous learning online presents several challenges to both instructors and students, yet, it provides learning opportunities for instructors to rethink how to bring forward some of the lessons from online learning back to physical classrooms. This is taken particularly in view of the various technological advancements and societal shifts that, ironically, enables us to transit from offline to online learning with minimum disruption in such trying times.

**Useful Lessons for the Physical Classes**

Bringing these insights back to a physical classroom environment, the question becomes how we may incorporate these online elements as an ally to teachers. While this assumes the widespread availability of physical mobile and laptop devices for both teachers and students, we believe that both the pandemic and the lowering of costs of such devices will drive the necessity and affordability of such set ups. There are two key elements which we propose are important in such hybrid set-up:

1. Online chatroom or two-way communication mechanism where both the teachers and students can see the student discussion(s) on-going virtually;
2. Anonymous response systems where collective feedback may be solicited, such as a poll, word cloud, multiple choices responses.

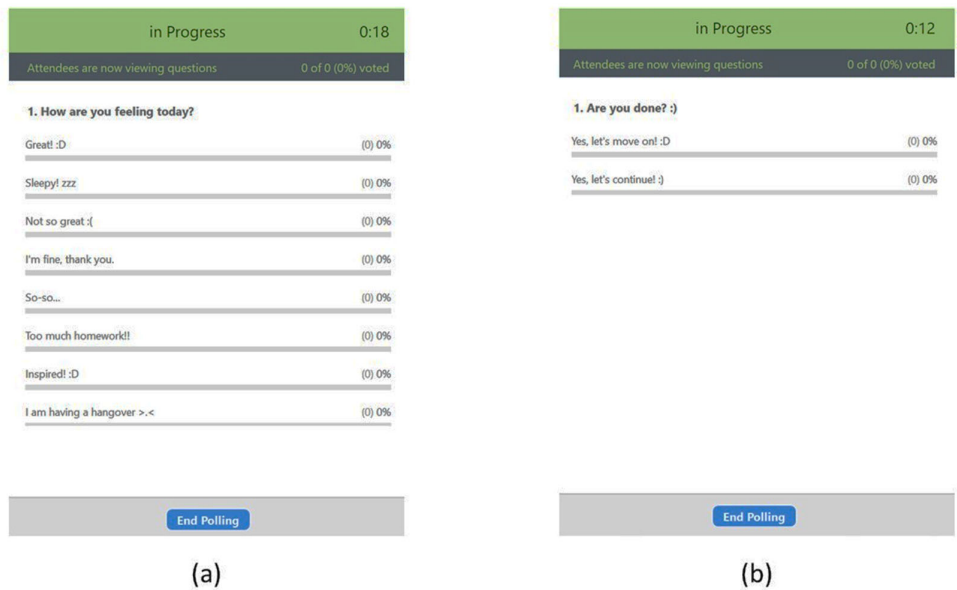

Figure 4 Two enactment of the use polls in the online classroom: (a) Polling the students' mental state at the start of the class. This checks on the students' well-being. While students' responses are anonymous, instructors can highlight that they are available for help should there be any negative responses from the students. This is especially important given that students are learning remotely, and this can bring some sense of inclusivity to the students. (b) A simple question to check if they are done with the task. Unlike a regular class, teachers are not able to walk around to check on students' progress, so active feedbacks like this from the students will be necessary.

A key takeaway from our full synchronous online learning experience is that such platform creates additional affordance for students' engagement within the virtual classroom, as seen in the following ways. The video-conference platform we used provides multiple ways of soliciting real-time feedback, such as instant online polls and live chat (not unlike that seen in on-line gaming platforms). We find that these additional avenues have encouraged students to express their misconceptions and misunderstanding, without the fear of being perceived as asking a trivial question or disrupting the flow of the class. This creates an enabler for even the least articulate students to express themselves. (Even 'shy' students can also use the "private chat" function.) We argue that given all students can "see" and follow the happenings in the chatroom, once questions are answered within the chatroom, students will in fact be encouraged to communicate because of the positive reinforcement. In our experience, the poll function has also been useful for multiple purposes, such as keeping track of students' progress during in-class activity, students' general well-being at the start of the class[21], solicit responses for in-class concept questions. Given that some of the video conferencing platforms are free to use and already contain these helpful features, we suggest running a parallel session deploying such platforms concurrently during physical classes, such that teachers can harness the advantages mentioned above. Such parallel session format will be trialed in our upcoming rerun of the course in the Fall term.

**Conclusion**

While the onset of the global pandemic led to the sudden transition of lessons from physical to online, and that much of these transitions would be inevitable reactive in nature, yet this creates an opportunity for a potential paradigm shift in the use of educational technologies in both the online and physical physics classroom.

**Acknowledgements**

We would like to thank our colleague, Wei Lek Kwan, for granting us the permission to use the left photo in Figure 1.

---

[1] McBrien, J. L., Cheng, R., & Jones, P. (2009). Virtual spaces: Employing a synchronous online classroom to facilitate student engagement in online learning. International review of research in open and distributed learning, 10(3).

[2] Fadde, P. J., & Vu, P. (2014). Blended online learning: Benefits, challenges, and misconceptions. Online learning: Common misconceptions, benefits and challenges, 33-48.

[3] Gray, J., & Lindstrøm, C. (2019). Five Tips for Integrating Khan Academy in Your Course. The Physics Teacher, 57(6), 406-408.

[4] Korsunsky, B., & Li, C. (2017). By hook or by MOOC: Lessons learned and the road ahead. The Physics Teacher, 55(3), 146-148.

[5] Hansch, A., Hillers, L., McConachie, K., Newman, C., Schildhauer, T., & Schmidt, J. P. (2015). Video and online learning: Critical reflections and findings from the field.

[6] Thomas, J. (2020). Positives from Online Teaching. Physics, 13(130)

[7] Rai, B. (2020). A team of instructors' response to remote learning due to Covid-19. A 10.012 Introduction to Biology case study. Journal of Applied Learning and Teaching, 3(2).

[8] Guo, S. (2020). Synchronous versus asynchronous online teaching of physics during the COVID-19 pandemic. Physics Education, 55(6), 065007.

[9] Pols, F. (2020). A Physics Lab Course in Times of COVID-19. Electronic Journal for Research in Science & Mathematics Education, 24(2), 172-178.

[10] https://smt.sutd.edu.sg/education/undergraduate/courses/physical-world

[11] Usually there are about 8 to 10 classes running concurrently during the regular term.

[12] There are other challenges that we have omitted in this article for brevity and focus: As part of the coursework, students were usually tasked to engage in a term-long project where they would build a physical prototype to tackle the given challenge, incorporating and considering the physics concepts that they learnt within the class in their prototype and solution. Careful design would be needed for such collaborations between students to take place in a pure online environment. Secondly, the formal assessment would have to be modified due to the inability to conduct physical assessments.

[13] This is in contrast with physical classrooms where bigger demonstration kits (in terms of dimensions) are desired, so that the entire class can see.

[14] At least in situations where students are in an environment that permits their web camera to be turned on, or if they have a web camera. Most modern laptops have built in cameras, but for students who uses a desktop, this may involves purchasing an additional external web camera, In that case, we no longer insist students keep their cameras on. On this matter, we give the benefit of doubt to students as much as possible. Interestingly, only one of our students who had the camera off remained uncontactable and did not take part in most of the class activities (this was evident from the fact that the student did not enter his assigned breakout rooms).

[15] However, we do acknowledge that this is subjected to the affordance of the resources available to the instructors.

[16] We would argue that this is easier to be done in the full online context, given that students are all learning remotely. Hence, the traditional role of teachers being a disciplinarian is drastically reduced in such context.

[20] This article does not seek to endorse any product and hence the platform we used has been omitted. However, we have tried various products and find that most of the products available at the time of writing can fulfil the 2 elements in the main text that we think are the most important. The decision matrix on the suitability of the product goes beyond the scope of this article.

[21] The idea of anonymity is crucial in our view here. Students are more likely to provide truthful responses if they realize that they will not be tracked. In practice, what we will do is then that announce to the students to come look for us in private if they have indicated that they have issues.